\documentclass[sigconf]{acmart}
\usepackage{siunitx}
\usepackage{graphicx}
\usepackage{float}
\usepackage{caption}
\usepackage{subcaption}
\usepackage{comment}
\usepackage{mathtools, multirow}
\usepackage{color}
\usepackage{pgfplots}
\pgfplotsset{compat=1.18} 

\renewcommand\footnotetextcopyrightpermission[1]{}

\usepackage{tikz}
\usepackage{xcolor}
\usetikzlibrary{arrows.meta, positioning}

\tikzset{
  box/.style={draw, rounded corners, minimum width=1.5cm, minimum height=1.2cm, align=center},
  optical/.style={box, fill=blue!15, draw=black!200},
  electronic/.style={box, fill=green!20, draw=black!200},
  cache/.style={box, fill=green!15, draw=black!100},
  smallbox/.style={box, fill=gray!15, draw=black!200},
  grad/.style={box, fill=green!20, draw=black!200},
  sparse/.style={box, fill=teal!10, draw=black!200},
  opt/.style={box, fill=red!20, draw=black!200}
}

\usepackage{multirow, booktabs, graphicx}
\usepackage{placeins, xcolor, amsmath}

\AtBeginDocument{%
  \providecommand\BibTeX{{%
    \normalfont B\kern-0.5em{\scshape i\kern-0.25em b}\kern-0.8em\TeX}}}

\begin{document}
\pagestyle{plain}

\title{FrequencyFormer: A Co-Designed Sensor-to-Processor Pipeline for Frequency-Domain Vision Transformer Inference}

\author{
Chengwei Zhou$^{1}$, Ovishake Sen$^{2}$, Xuming Chen$^{1}$, Rishith Paramasivam$^{2}$, Shaahin Angizi$^{3}$,\\
Swarup Bhunia$^{2}$, Baibhab Chatterjee$^{2}$, and Gourav Datta$^{1}$  \\
\vspace{0.8mm}
\small
$^{1}$Case Western Reserve University, USA \quad
$^{2}$University of Florida, USA \quad
$^{3}$New Jersey Institute of Technology, USA
{\tt\small \{chengwei.zhou,xuming.chen,gourav.datta\}@case.edu}\quad
{\tt\small \{ovishake.sen,rparamasivam,swarup,chatterjee.b\}@ufl.edu}\quad
{\tt\small shaahin.angizi@njit.edu}
\\}






\begin{abstract}
The deployment of vision transformers (ViTs) on sensor-edge systems is fundamentally bottlenecked not by on-device computation alone, but by the energy and bandwidth cost of transmitting high-dimensional image data from sensor to processor. In-sensor and near-sensor computing paradigms aim to address this by performing early feature extraction at the sensor, yet existing approaches achieve only modest compression ratios, limiting their practical impact. We observe that the frequency domain offers a natural and highly compact representation of visual information, and that this compactness can be exploited at the sensor level to dramatically reduce the volume of data that must traverse the sensor-processor interface. Building on this insight, we present FrequencyFormer, a co-designed sensor-to-processor pipeline for efficient ViT inference comprising three tightly integrated components: (1) a multi-scale discrete cosine transform (DCT) tokenizer that compresses a 224x224 input image into a compact set of frequency-domain tokens, achieving up to 128x reduction in off-chip data volume with modest accuracy loss on standard benchmarks; (2) a look-up table (LUT)-based hardware realization of the tokenizer on a near-sensor logic chip, leveraging the fixed-coefficient structure of DCT to enable precomputed, energy-efficient, and area-efficient inference without multipliers; and (3) a modified Mobile Industry Processor Interface (MIPI)-based low-power communication architecture that further reduces energy per bit of the compressed data transfer, yielding multiplicative savings on top of the bandwidth reduction. FrequencyFormer operates as a drop-in replacement for the standard ViT patch embedding, preserving compatibility with pretrained backbones across classification, detection, and segmentation tasks. FrequencyFormer achieves 28.8 TOPS/W, while maintaining accuracy. The full pipeline reduces communication energy by $\approx$230$\times$ and total sensor-side energy by 2.22$\times$ compared to conventional architectures, establishing frequency-domain tokenization as a viable foundation for scalable in-sensor ViT deployment.
\end{abstract}





\maketitle

\section{Introduction}

\begin{figure}[t]
  \centering
\includegraphics[width=\linewidth]{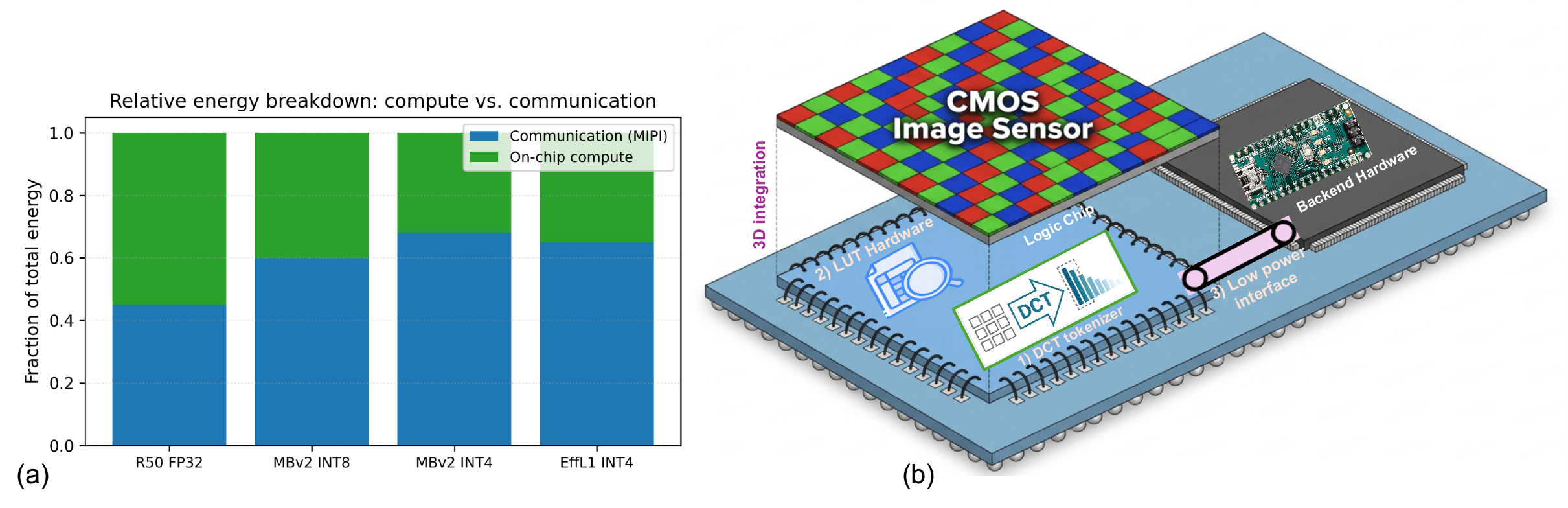}
  \vspace{-4mm}
\caption{\footnotesize (a) Energy breakdown for edge vision backbones on 360×360 frames shows that as models are optimized (ResNet-50 FP32 → MobileNet-v2 INT8/INT4 → EfficientFormer-L1 INT4), compute energy decreases while communication (MIPI) remains constant, increasing its share of total energy. Estimates use ASIC-level models including compute and memory.
(b) Proposed 3D-integrated pipeline with a stacked CMOS sensor and near-sensor logic performing LUT-based DCT tokenization, transmitting compact tokens via a low-power interface.}
  \label{fig:freq_framework}
  \vspace{-2mm}
\end{figure}

Vision transformers (ViT) have become the dominant architecture for high-performance image understanding, achieving state-of-the-art results across classification, detection, and segmentation tasks~\cite{dosovitskiy2020image, touvron2021training}. Their success, however, comes at the cost of substantial computational and memory requirements, making deployment on resource-constrained edge platforms challenging. Recent work has focused on reducing the compute of transformer inference, via quantization~\cite{jacob2018quantization, vasquez2021activation}, pruning~\cite{han2016deep, liang2021pruning}, and token reduction~\cite{rao2021dynamicvit, bolya2023token}, but a critical and often underappreciated bottleneck remains: the energy and bandwidth cost of moving raw image data from the sensor to the processor. In a typical edge vision system, the image sensor and the inference engine reside on separate chips, so the full-resolution frame must be transmitted across an off-chip interface before any computation can begin. For a standard $360\times360$ RGB image at 8-bit precision, this corresponds to roughly 3.1 megabits per frame. At the data rates and frame rates required by real-time applications, the energy spent on this communication link can rival or even exceed the energy of the inference itself, particularly once the backbone has been aggressively optimized for edge deployment, as illustrated in Fig.~\ref{fig:freq_framework}(a).
 
The paradigm of in-/near-sensor computing has emerged as a response to this challenge~\cite{mennel2020ultrafast, zhou2020near, chai2023insensor}. Rather than transmitting the full image and processing it remotely, in-sensor computing performs early-stage neural network operations directly at the pixel array or on a co-integrated logic chip adjacent to the sensor, transmitting only a compressed intermediate representation to the downstream processor. Recent work has demonstrated in-sensor implementations of simple multi-layer perceptron as well as convolutional layers within pixel arrays~\cite{adibi2022insensor, datta2022processing}. While these efforts represent meaningful progress, they share a common limitation: the compression ratios they achieve are modest, typically ranging up to $10\times$. At these levels, the bandwidth savings may not justify the added complexity and silicon cost of integrating compute logic with or near the sensor. For in-sensor computing to deliver on its promise, the compression achieved at the sensor must be dramatic enough to fundamentally alter the system energy balance.
 
We argue that the frequency domain is a natural representation for addressing sensor–processor bottlenecks. The discrete cosine transform (DCT) concentrates most visual energy into a small set of low-frequency coefficients—a property long exploited in JPEG~\cite{wallace1992jpeg} and validated for machine learning, where sparse DCT inputs achieve competitive accuracy~\cite{xu2020learning,gueguen2018faster}. Unlike learned compression (e.g., autoencoders), DCT provides aggressive dimensionality reduction via a fixed, lightweight linear transform with no learned parameters or iterative encoding, making it well-suited for near-sensor implementation under strict area and power constraints. Its fixed basis also enables hardware specialization at fabrication time. Building on this insight, we propose \textit{FrequencyFormer}, a co-designed sensor-to-processor pipeline comprising three tightly coupled components as shown in Fig.~\ref{fig:freq_framework}(b).

\noindent\textbf{Multi-scale DCT tokenizer.} The first component decomposes the input image (YCbCr) at three spatial scales ($8{\times}8$, $32{\times}32$, and global $224{\times}224$) using block-wise DCT, followed by channel selection to retain the most informative frequency coefficients. The multi-resolution features are fused into a compact token sequence for standard ViTs, achieving up to $128\times$ data reduction. Since channel indices are fixed after training (via Gumbel-Softmax~\cite{jang2017categorical} or zigzag ordering), inference uses only fixed-coefficient operations.

\noindent\textbf{LUT-based hardware realization.} This fixed structure enables efficient LUT-based acceleration. DCT coefficients are constants, allowing coefficient-pixel products to be hardwired into ROM-backed LUTs, eliminating multipliers and SRAM overhead. We adopt a divide-and-conquer LUT design~\cite{sen2025look} and further reduce cost via mixed precision: low-frequency components use INT8, while high-frequency components use INT4. Learned weights are stored in a smaller SRAM-backed LUT datapath, yielding significant energy and area savings.


\noindent\textbf{Low-power communication interface.} The communication link is co-designed with the compressed representation. The tokenizer reduces data volume by $128\times$, while a modified MIPI interface achieves an additional $1.8\times$ per-bit energy reduction. We adopt an asymmetric architecture that shifts complexity to the receiver (FPGA/custom), which employs integrating detection to achieve low BER, while keeping the sensor-side lightweight and energy-efficient.


Together, these three components form a complete pipeline from sensor to processor. The image sensor captures the scene; the near-sensor logic chip performs DCT decomposition, channel selection, convolution, and optional cross-attention fusion using LUT-based hardware; the compressed token stream is transmitted over the optimized communication link; and the main processor runs the standard ViT backbone on the received tokens.  
 
 We evaluate FrequencyFormer on image classification (CIFAR-10, Tiny-ImageNet, VWW) and dense prediction (object detection and instance segmentation on COCO~\cite{lin2014coco}), using ViT-Tiny, ViT-Base, Swin-Tiny, and EfficientFormer-L1 backbones. Our experiments demonstrate that the multi-scale DCT tokenizer preserves near-baseline accuracy across all tasks while reducing off-chip data volume by up to $128\times$. The LUT-based realization with HAQ achieves 28.8 TOPS/W---2.5$\times$ higher energy efficiency than prior in-/near-sensor work. The combined pipeline reduces communication energy by $\approx$230$\times$ and total sensor-side energy by 2.22$\times$ relative to a conventional baseline.

\section{Related Work}

\noindent\textbf{Frequency-Domain Representations for Visual Recognition}: Frequency-domain representations (e.g., DCT) compactly encode images by concentrating energy in a few coefficients~\cite{wallace1992jpeg}. Prior work shows neural networks can operate directly on such representations, including frequency-domain CNNs~\cite{chen2016packing,gueguen2018faster} and ImageNet-scale models with reduced input dimensionality~\cite{ehrlich2019deep,xu2020learning}. Recent transformer-based approaches incorporate spectral processing—e.g., replacing MHSA with Fourier mixing (FNet~\cite{lee2022fnet}) or integrating frequency components into attention (FcaNet~\cite{qin2021fcanet}, SpectFormer~\cite{patro2025spectformer}), as well as DCT-based tokenization and attention compression~\cite{li2023dcformer,pan2024dctattention,lee2024dctvit}. While learned compression methods~\cite{balle2017end,balle2018variational} and token reduction techniques (DynamicViT~\cite{rao2021dynamicvit}, ToMe~\cite{bolya2023token}) reduce compute, they either require significant preprocessing or operate after full image transmission, limiting their applicability at the sensor.

\noindent\textbf{LUT Acceleration}:
LUT-based neural acceleration has been explored to improve energy efficiency by replacing conventional MAC operations with precomputed memory accesses. This has been shown in the context of systolic arrays ~\cite{liu2022lut}, in-cache processing framework for neural workloads ~\cite{ramanathan2020look}, programmable compute-in-memory (CiM) fabric ~\cite{dehghanzadeh2025luna}, as well as scalable neural accelerators using a divide-and-conquer (D\&C) strategy ~\cite{sen2025look}, achieving significant reductions in power and power-area product in each scenario. 

\noindent\textbf{Low-power Communication Interface}: Recent work on ultra-low-power camera-to-processor communication has moved from conventional high-speed sensor readout toward co-designed sensing, compression, and interface architectures that minimize the number of bits leaving the focal plane \cite{GT_Comp}. Both electrical and optical domain processing with communication \cite{photonic_in-sensor, ISSCC21_Comm, JSSC14_Comm} results in roughly 1-20 pJ-per-pixel cost for state-of-the-art sensing/blind-compression pipelines, with $>$1 nJ-per-pixel for communicating the data. However, most of these works focus on optimizing computation-communication trade-offs, without providing much insight on if the communication link itself can be made much lower-power.

\section{Proposed Method}
\begin{figure}[t]
  \centering
\includegraphics[trim={85 192 100 170}, clip, width=\linewidth]{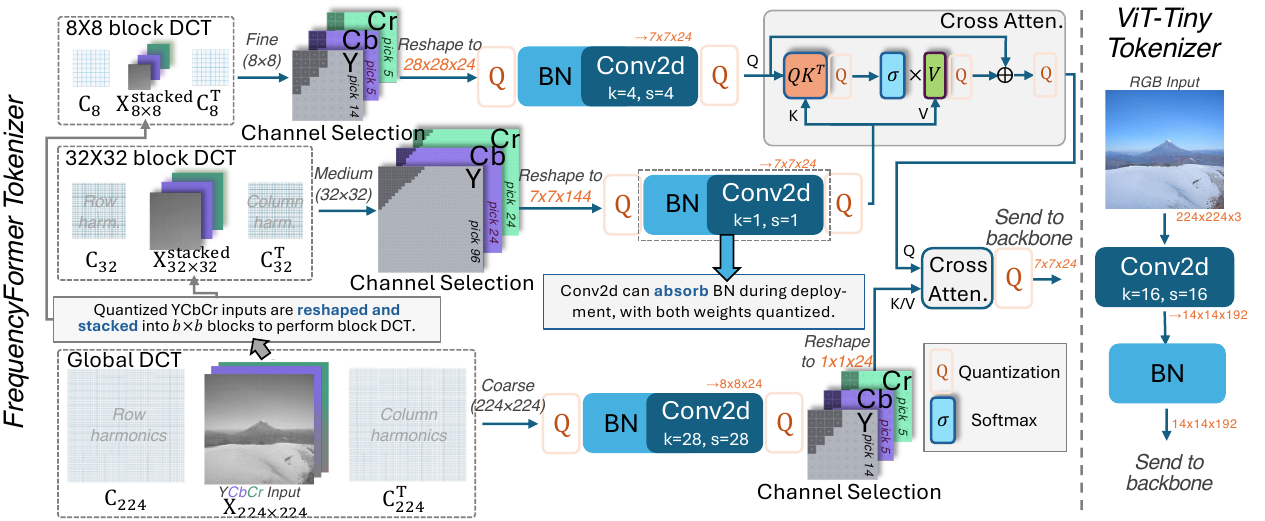}
  \vspace{-2mm}
\caption{\footnotesize Overview of the FrequencyFormer tokenizer. The input YCbCr image is decomposed via multi-scale block-wise and global DCT. At each scale, selected frequency coefficients are quantized and projected through BN-fused strided convolutions into compact spatial tokens. A cascaded cross-attention module fuses the three branches before forwarding to the backbone. All weights and activations are integer-quantized. Feature dimensions in \textcolor{orange}{orange} correspond to the $7{\times}7$, 24-channel output configuration.}
  \label{fig:tokenizer}
  \vspace{-4mm}
\end{figure}

\begin{figure}[t]
  \centering
\includegraphics[trim={150 235 160 220}, clip, width=\linewidth]{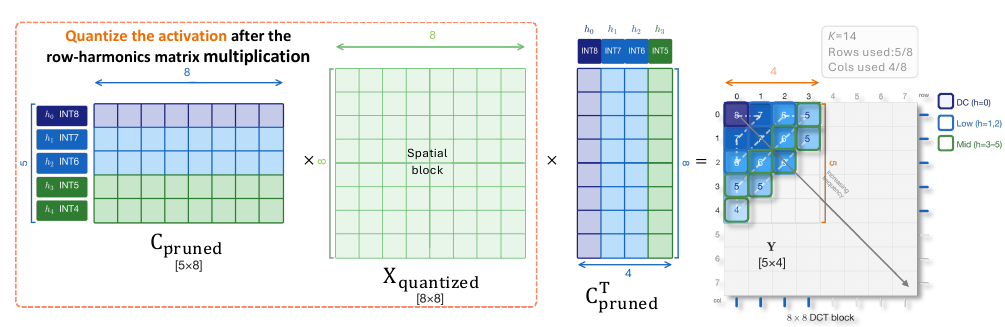}
  \vspace{-6mm}
\caption{\footnotesize 8$\times$8 Selection-Aware DCT Pruning with Harmonic-Aware Quantization. $\mathbf{C_8}$ is pruned to 5 rows and $\mathbf{C_8}^{\!\top}$ to 4 columns needed by the $K{=}14$ zigzag coefficients, reducing to a $5{\times}4$ output tile. Bit-widths decrease from INT8 to INT4 per row/column; the intermediate result $\mathbf{C}_{\mathrm{pruned}}\mathbf{X}$ is requantized before the column-harmonic multiplication. The resulting $5\times4$ coefficient matrix $\mathbf{Y}$, where each cell shows its effective precision (min of row and column bit-widths), overlaid with the 
zigzag scan order and frequency-band annotations.}
  \label{fig:DCT}
  \vspace{-4mm}
\end{figure}
 
FrequencyFormer is a three-component sensor-to-processor pipeline for ViT inference in bandwidth-constrained systems. The pipeline partitions the inference task across two physical domains: a near-sensor logic chip and an off-chip main processor. On the near-sensor chip, a multi-scale DCT tokenizer (Section~\ref{sec:tokenizer}) transforms the raw image into a compact frequency-domain token sequence, implemented using energy-efficient LUT-based hardware (Section~\ref{sec:lut}). The compressed tokens are then transmitted to the main processor over a low-power communication interface (Section~\ref{sec:comm}), where a standard ViT backbone performs classification or dense prediction on the received tokens. The architectural split is governed by a simple principle: operations whose coefficients are fixed or can be hardened after training belong on the sensor chip, where they benefit from LUT-based precomputation; the remaining learned transformer layers, which require programmable and higher-precision computation, execute on the main processor. 

 
\subsection{Multi-Scale Frequency-Domain Tokenizer}
\label{sec:tokenizer}
 
The tokenizer converts a $224{\times}224$ RGB image into a short
sequence of frequency tokens through four components (as shown in Fig.~\ref{fig:tokenizer}). All weights and activations are represented in integer precision to enable efficient LUT-based precomputation, while softmax remains in FP16. 
Each components addresses a specific bottleneck:
(i)~multi-scale DCT decomposes the image into complementary
    spatial-frequency features (Section~\ref{sec:msdct});
(ii)~Selection-Aware DCT Pruning (Section~\ref{sec:pruning});
(iii)~harmonic-aware quantization allocates precision
    proportional to spectral importance
    (Section~\ref{sec:quant}); and
(iv)~multi-branch fusion merges the multi-scale
    representations into a unified token sequence
    (Section~\ref{sec:fusion}).

 
 
\subsubsection{Multi-Scale DCT Decomposition}
\label{sec:msdct}

The input is first converted to YCbCr color space; following
standard practice~\cite{wallace1992jpeg, ibraheem2012ycbcr}.
For an $N{\times}N$ spatial block $\mathbf{X}$, the 2-D DCT
computes the coefficient matrix
\begin{equation}
\vspace{-1mm}
  \mathbf{Y} = \mathbf{C}\,\mathbf{X}\,\mathbf{C}^{\!\top},
  \label{eq:dct2d}
\end{equation}
where $\mathbf{C}\!\in\!\mathbb{R}^{N\times N}$ is the orthonormal
DCT basis with entries
\begin{equation}
\vspace{-1mm}
  C_{k,i}
  = w_k\,\cos \Bigl(\tfrac{\pi(2i{+}1)k}{2N}\Bigr),
  \quad
  w_k =
  \begin{cases}
    \sqrt{1/N} & k=0,\\
    \sqrt{2/N} & k>0.
  \end{cases}
  \label{eq:dct_basis}
  \vspace{-1mm}
\end{equation}
The tokenizer processes YCbCr through three parallel branches at different block sizes to capture complementary spatial-frequency features. 

\noindent\textit{Branch 1 (Local features, $8\times8$ blocks).} The Y, Cb, and Cr channels are each partitioned into non-overlapping $8\times8$ blocks, and the 2D DCT is applied independently to each block. This block size is chosen for compatibility with the JPEG standard and captures fine-grained local texture and edge information. For a $H\times{W}$ single-channel image, $8\times8$ block-wise DCT produces a grid of $\frac{H}{8}\times\frac{W}{8}$ blocks, each containing 64 frequency coefficients. After channel selection (described below), we retain $k_y$, $k_{cb}$, and $k_{cr}$ coefficients from Y, Cb, and Cr respectively, yielding a feature map of size $\frac{H}{8}{\times}\frac{W}{8}{\times}(k_y{+}k_{cb}{+}k_{cr})$. A 2D convolution with kernel size $k$ and stride $s$ then reduces the spatial resolution to $\frac{H}{8{\cdot}s}\times\frac{W}{8{\cdot}s}$, producing a token map of shape $\frac{H}{8{\cdot}s}\times\frac{W}{8{\cdot}s}\times(k_y{+}k_{cb}{+}k_{cr})$.
 
\noindent\textit{Branch 2 (Medium-scale structure, $32\times32$ blocks).} The same YCbCr channels are partitioned into $32\times32$ blocks and transformed via block-wise DCT. The larger block size captures medium-scale structural patterns such as object boundaries and semantic regions that span multiple $8\times8$ blocks but remain local relative to the full image. After channel selection, the resulting feature map is passed through batch normalization followed by a 2D convolution (kernel size 1, stride 1) to project the selected coefficients into a 24- or 48-channel embedding, depending on the channel configuration described in Section~\ref{exp_set}. This produces tokens of shape $\frac{H}{32}\times\frac{W}{32}\times 24/48$.
 
\noindent\textit{Branch 3 (Global semantics, full-image DCT).} Unlike the first two branches, Branch 3 applies the 2D DCT globally to the $H\times{W}$ image for each of the Y, Cb, and Cr channels, capturing the global frequency structure including overall illumination, color distribution, and large-scale layout. The resulting $H\times{W}\times3$ coefficient map is processed through a batch normalization and a 2D convolution layer with kernel size 28 and stride 28, which performs a form of learned pooling in the frequency domain, reducing the spatial extent to $\frac{H}{28}\times\frac{W}{28}\times3$. Channel selection then retains the most informative global coefficients ($k_y$ from Y, $k_{cb}$from Cb, and $k_{cr}$ from Cr), producing a token map of shape $1\times1\times(k_y{+}k_{cb}{+}k_{cr})$.

The three branches are deliberately designed to span complementary frequency features. Branch 1 captures high-to-mid frequency local detail (textures, edges, fine patterns). Branch 2 captures mid-frequency structural information (object contours, part boundaries). Branch 3 captures low-frequency global information (scene layout, illumination gradients). This multi-scale decomposition in the frequency domain is analogous to the spatial multi-scale feature pyramids used in convolutional architectures, but operates on frequency coefficients rather than pixel activations.

Within each branch, a subset of $K$ DCT coefficients is retained via the JPEG zigzag scan, which traverses the coefficient matrix from low to high frequency along diagonal paths, providing an effective fixed selection priority without learnable parameters.
As an alternative, we evaluate learnable selection via Gumbel-Softmax~\cite{jang2017categorical}, where a soft mask $m_i = \sigma\bigl((g_i + \log\alpha_i)/\tau\bigr)$ is applied to each coefficient, with $g_i$ drawn from a Gumbel distribution, $\alpha_i$ the learnable logit, and $\tau$ the temperature. The masks are jointly optimized during training and hardened to fixed indices at inference. Although this yields up to 0.2\% accuracy improvement, the resulting non-contiguous coefficient positions prevent the row/column pruning mentioned in Section~\ref{sec:pruning} , negating computational savings. We therefore adopt zigzag ordering as the default selection strategy.
\textit{In both cases, the selection reduces to address routing at inference
and incurs {zero} additional computation.}

\subsubsection{Selection-Aware DCT Pruning}
\label{sec:pruning}
 
Regardless of how the $K$ retained coefficients are chosen (via
zigzag ordering or learned selection), each selected coefficient
occupies a position $(u, v)$ in the 2-D frequency grid and
therefore requires only row~$u$ of $\mathbf{C}$ and column~$v$ of
$\mathbf{C}^{\!\top}$.
Collecting the distinct row indices into $\mathcal{R}$ and the
distinct column indices into $\mathcal{S}$ across all $K$ positions,
we have $R=|\mathcal{R}|$ rows and $S=|\mathcal{S}|$ columns that
are actually needed.
We prune $\mathbf{C}$ in Eq. 1 to retain these $R$ rows, and $\mathbf{C}^{\top}$ to retain these $S$ columns, yielding:
\begin{equation}
  \mathbf{Y}_{[K]}
  = \mathbf{C}_{\mathrm{row}}\,\mathbf{X}\,
    \mathbf{C}_{\mathrm{col}}^{\!\top},
  \qquad
  \mathbf{C}_{\mathrm{row}} \in \mathbb{R}^{R \times N},\;
  \mathbf{C}_{\mathrm{col}} \in \mathbb{R}^{S \times N},
  \label{eq:pruned_dct}
  \vspace{-2mm}
\end{equation}
reducing the complexity from $\mathcal{O}(N^{3})$ to
$\mathcal{O}((R{+}S){\cdot}N^2)$, with the output being an
$R{\times}S$ tile from which the $K$ coefficients are read directly.
For $N{=}8$ and $K{=}14$, the selected positions occupy $R{=}5$
rows and $S{=}4$ columns (Fig.~\ref{fig:DCT}), producing a compact
$5{\times}4$ output.

\subsubsection{Harmonic-Aware Quantization(HAQ)}
\label{sec:quant}
 
Uniform quantization across all harmonic rows ignores their
energy hierarchy.
Empirically, we find that aggressively quantizing high-frequency
harmonics (e.g., to 4-bit) causes negligible accuracy loss, confirming that the semantic content of DCT representations is primarily concentrated in
the low-frequency components.
Motivated by this observation, we assign \emph{linearly
decreasing} bit-widths to the rows of $\mathbf{C}_{\mathrm{row}}$
according to harmonic order~$k$:
\begin{equation}
  b_k = \mathrm{round}\!\left(
    b_{\max} - \frac{(b_{\max}-b_{\min})\,k}{t}
  \right),
  \quad k = 0,\ldots,R{-}1,
  \label{eq:bit_schedule}
\end{equation}
with $b_k = b_{\min}$ for $k > t$, where $b_{\max}$ is maximum bit width,
$b_{\min}$ is the minimum bit width, and $t$ is a transition harmonic index.
Each row is independently quantized via symmetric uniform
quantization:
\begin{equation}
  \widetilde{C}_{k,\cdot}
  = \Delta_k \cdot \mathrm{clip}\!\left(
      \left\lfloor \frac{C_{k,\cdot}}{\Delta_k}\right\rceil,\;
      -2^{b_k-1},\; 2^{b_k-1}{-}1
    \right),
  \quad
  \Delta_k = \frac{\|C_{k,\cdot}\|_\infty}{2^{b_k-1}-1}.
  \label{eq:row_quant}
  \vspace{-2mm}
\end{equation}

\noindent\textit{Matched-precision multiplication via bit-truncation.}
To fully exploit the reduced bit-widths, we enforce matched-precision arithmetic: the $k$-th harmonic row $\widetilde{C}_{k,\cdot}$ at $b_k$ bits is multiplied with $\mathbf{X}$ also at $b_k$ bits, so the MAC unit operates as INT$b_k{\times}$INT$b_k$ (e.g., INT8${\times}$INT8 for DC, INT4${\times}$INT4 for high-frequency harmonics), enabling narrower, more energy-efficient multipliers for high-frequency rows. This does not require storing multiple copies of $\mathbf{X}$ at different precisions: since $\mathbf{X}$ is already quantized to INT$b_{\max}$, each row simply reads the upper $b_k$ bits of the stored representation, with the effective scale factor $\Delta^{x}_{k} = \Delta^{x}\!\cdot 2^{\,b_{\max}-b_k}$ absorbed as a compile-time constant into the dequantization step. In hardware, this amounts with zero additional storage or arithmetic overhead.


The intermediate result $I_{k,\cdot} = \widetilde{C}_{k,\cdot}\;X_{\mathrm{int}}^{(b_k)}$ (INT$b_k{\times}$INT$b_k$) is re-quantized to $b_k$ bits before the column-harmonic multiplication $\mathbf{Y}_{[K]} = \widetilde{\mathbf{I}}\,\widetilde{\mathbf{C}}_{\mathrm{col}}^{\!\top}$, keeping the entire two-stage computation at matched low precision. Truncation error is incurred only for high-frequency harmonics whose basis vectors have small magnitude, so the product error is suppressed by the decaying spectral energy; DC and low-frequency rows use $b_k = b_{\max}$ and lose no precision. This design aligns hardware cost with spectral importance (Fig.~\ref{fig:DCT}, left).

\subsubsection{Multi-Branch Fusion}
 \label{sec:fusion}
 
The outputs of the three branches must be combined into a unified token sequence for the downstream ViT backbone. We propose a sequential cross-attention fusion that hierarchically distills information across frequency scales. Let $\mathbf{T}_1$, $\mathbf{T}_2$, and $\mathbf{T}_3$ denote the token outputs of Branches 1, 2, and 3, respectively. The fusion proceeds in two successive cross-attention stages~\cite{vaswani2017attention}. In the first stage, $\mathbf{T}_1$ serves as the query and $\mathbf{T}_2$ provides the keys and values:
\begin{equation}
\mathbf{T}_{12} = \texttt{softmax}\left(\frac{\mathbf{T}_1 \mathbf{W}_Q^{(1)} \left(\mathbf{T}_2 \mathbf{W}_K^{(1)}\right)^\top}{\sqrt{d_k}}\right) \mathbf{T}_2 \mathbf{W}_V^{(1)},
\label{eq:xattn1}
\vspace{-2mm}
\end{equation}
allowing the local fine-grained tokens of Branch 1 to attend over the medium-scale structural tokens of Branch 2. In the second stage, $\mathbf{T}_{12}$ queries against the global token $\mathbf{T}_3$:
\begin{equation}
\mathbf{T_{\texttt{out}} = \texttt{softmax}\left(\frac{\mathbf{T}_{12} \mathbf{W}_Q^{(2)} \left(\mathbf{T}_3 \mathbf{W}_K^{(2)}\right)^\top}{\sqrt{d_k}}\right) \mathbf{T}_3 \mathbf{W}_V^{(2)}},
\label{eq:xattn2}
\vspace{-2mm}
\end{equation}
modulating the fused local-structural representation with scene-level frequency context. The final output $\mathbf{T}_{\mathrm{out}}$ encodes all three frequency scales into a single compact token sequence, which is passed to the downstream backbone after a learned linear projection to the backbone embedding dimension $d$. Since the token counts and projection dimensions are small, the two cross-attention operations add negligible compute overhead relative to the bandwidth savings of the tokenizer.
 
 

\subsubsection{Integration with Vision Transformer Backbones}

FrequencyFormer is designed to be fully compatible with pretrained vision transformer ecosystem, as a drop-in replacement for the standard patch embedding layer, enabling backbones to be used without architectural changes. The frequency-domain tokens produced by the tokenizer are upsampled to match the input dimensions expected by the first transformer block. A lightweight linear projection maps the tokenizer output to the backbone embedding dimension (e.g., 192 for ViT-Tiny, 768 for ViT-Base), and interpolation-based spatial upsampling restores the standard token count. Standard positional embeddings are added, and the tokens are passed directly into the pretrained transformer blocks. Quantization-aware training (QAT) is employed to ensure compatibility with LUT-based deployment.

Importantly, the projection is performed on the processor side, after the compressed tokens are transmitted. The sensor–processor interface carries only the compact frequency representation, preserving the full bandwidth savings of the tokenizer.

To bridge the domain gap between projected frequency tokens and the backbone's spatial expectations, we introduce a tokenizer knowledge distillation (T-KD) loss: $\mathcal{L}_{\mathrm{T\text{-}KD}} = \left\| \mathbf{T}_{\mathrm{freq\_proj}} - \mathrm{sg}(\mathbf{T}_{\mathrm{vanilla}}) \right\|^2$, where the vanilla patch embedding is frozen as a reference and $\mathrm{sg}(\cdot)$ denotes the stop-gradient operator. The total loss is $\mathcal{L} = \mathcal{L}_{\mathrm{CE}} + \lambda\,\mathcal{L}_{\mathrm{T\text{-}KD}}$, with $\lambda$ controlling the distillation weight.

\begin{figure}[t]
  \centering
\includegraphics[width=\linewidth]{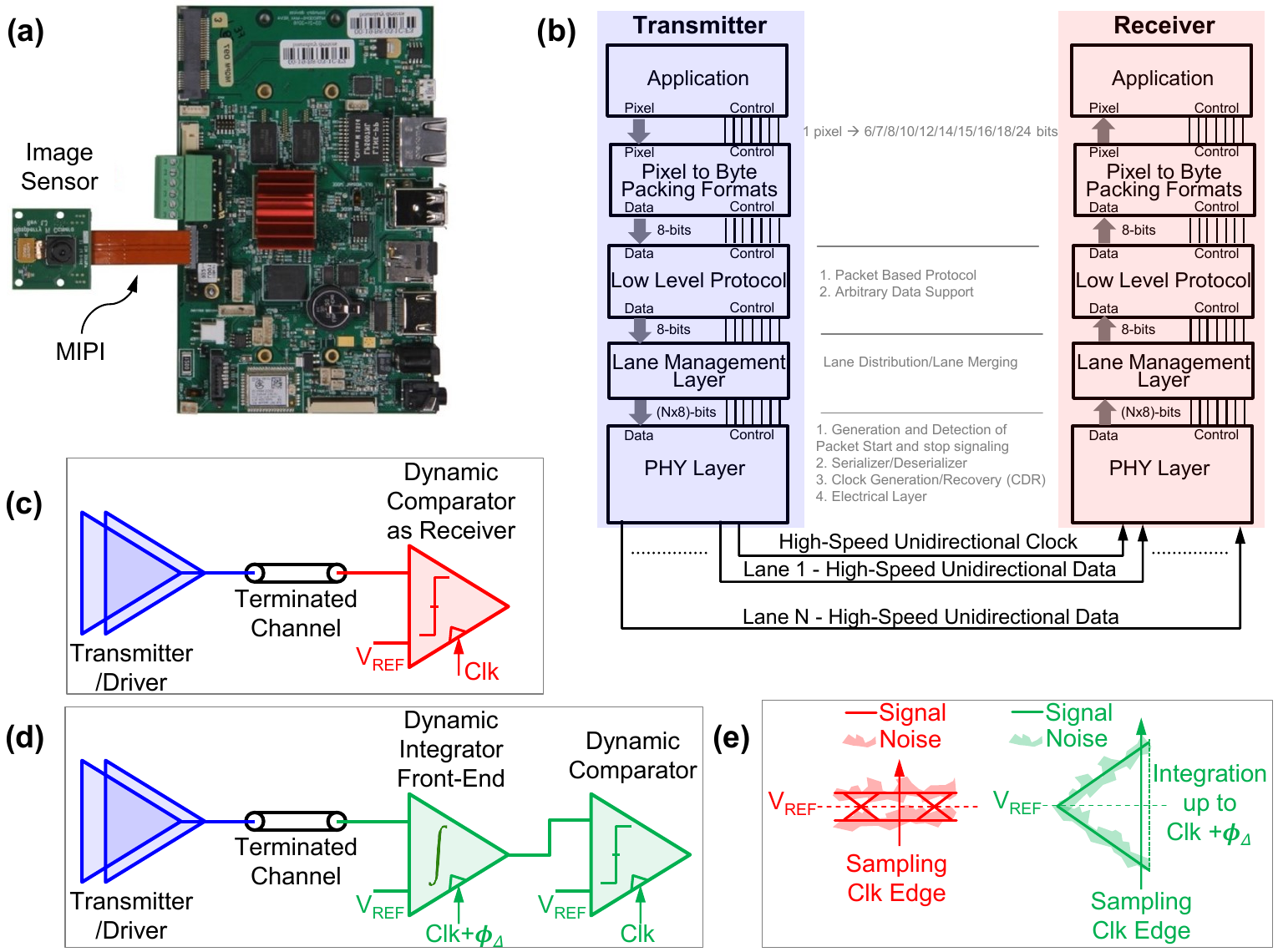}
  \vspace{-6mm}
\caption{\footnotesize (a) MIPI-CSI example; (b) Tx and Rx Architecture in MIPI \cite{MIPI} ; (c) Conventional Implementation of the Physical layer of the Tx/Rx, which is implemented with a simple inverter-based driver in the Tx and with a dynamic comparator at the Rx; (d) Proposed/Modified Implementation of the Physical layer of the Tx/Rx: the Tx remain unchanged, while the dynamic comparator at the Rx is preseded by a dynamic integrator; (e) eye-diagram representaion of the signal and noise for the scenario with just the dynamic comparator as the Rx (red), and for the scenario with the dynamic integrator+comparator as the Rx (green), showing an improved signal to noise ratio (SNR) for the integrator+comparator. This means that for iso-SNR scenario, the TX signal (and power) can be reduced for the integrator+comparator implementation.}
  \label{fig:Comm}
  \vspace{-6mm}
\end{figure}

\subsection{LUT-Based Hardware Realization}
\label{sec:lut}

Conventional multiplier-based MAC units dominate energy and area in neural accelerators. LUT-based computation replaces multiplications with precomputed memory lookups, reducing dynamic power and hardware complexity~\cite{sen2025look}. To address the exponential storage growth of traditional LUTs with bit precision, we adopt the divide-and-conquer (D\&C) strategy from~\cite{sen2025look}, which decomposes high-precision multiplications into low-precision sub-LUT operations combined via shift-and-add. We extend this framework beyond general-purpose acceleration by exploiting the specific structure of the FrequencyFormer tokenizer, whose compute decomposes into three operation classes with distinct LUT mapping characteristics.

\subsubsection{DCT Kernel Mapping}
The DCT basis matrix $\mathbf{C} \in \mathbb{R}^{N \times N}$ is defined entirely by the cosine function (Eq.~\ref{eq:dct_basis}) and is therefore mathematically fixed for a given block size. Unlike general DNN weights, which must be reprogrammed whenever the model changes, the DCT coefficients are universal constants: the same entries apply to every image, every task, and every deployment. This invariance has a direct hardware consequence. In a conventional LUT-based accelerator, the weight side of each multiplication is stored in SRAM and must be loaded from off-chip memory at initialization. For the DCT kernels, the LUT contents can instead be permanently hardwired into ROM or mask-programmed during fabrication, eliminating SRAM write circuitry, refresh logic, and the associated leakage power. For the $8 \times 8$ block-wise DCT with selection-aware pruning to $R{=}5$ rows, each row of $\mathbf{C}_{\mathrm{row}}$ contains $N{=}8$ fixed coefficients. Under D\&C with 2-bit sub-problems, each coefficient--input product decomposes into sub-LUTs of $2^2{=}4$ entries, requiring only 4 storage cells per sub-table. The total ROM footprint for the row-harmonic stage of a single $8 \times 8$ block is $R \times N \times 4 = 160$ entries, a negligible overhead that fits comfortably within the peripheral logic of a sensor die. Furthermore, Harmonic-Aware Quantization (Section~3.1.3) directly reduces the LUT hardware cost per harmonic row. High-frequency rows quantized to INT4 require only $4\mathrm{b} \times 4\mathrm{b}$ sub-LUTs (4 entries of 8-bit products), whereas DC and low-frequency rows at INT8 use $4\mathrm{b} \times 2\mathrm{b}$ sub-tables under D\&C decomposition. Since the number of high-frequency rows exceeds that of low-frequency rows in the pruned basis, the average LUT storage and multiplexer cost per block is substantially below the uniform full-precision case. This tight coupling between the spectral precision schedule and hardware cost is unique to the frequency-domain setting: in a general DNN layer, mixed precision must be determined via expensive sensitivity analysis, whereas in the DCT the energy hierarchy of harmonics provides a training-free guide for precision allocation.

\subsubsection{Convolution and Cross-Attention Mapping}
The strided 2D convolutions following channel selection involve learned weights that are fixed after training and quantized via QAT. These weights are programmed into SRAM-backed LUTs during deployment using the same D\&C decomposition as in~\cite{sen2025look}. Because the convolution kernels are small (e.g., $4 \times 4$ with 24 output channels for Branch~1), the total number of unique weight values is modest, and the SRAM cost is dominated by the DCT stages. Batch normalization parameters are folded into the convolution weights at deployment time, eliminating BN as a separate operation and further reducing the in-sensor compute footprint. The cross-attention projection weights 
$\mathbf{W}_Q, \mathbf{W}_K, \mathbf{W}_V$ are fixed after 
training, quantized, and stored in SRAM-backed LUTs; the 
input-dependent tokens serve as the data operands that 
index into these tables at runtime. The softmax nonlinearity is implemented in FP16 using a small dedicated functional unit, as it accounts for negligible area relative to the MAC-dominated projections. Since the token dimensions are compact ($49 \times 24$), the total number of MACs in the fusion stage is small (on the order of $10^4$), and the corresponding LUT area is a minor fraction of the overall in-sensor hardware budget.
 
\subsection{Low-Power Communication Interface}
\label{sec:comm}

We propose a low-power, asymmetric MIPI-based communication interface that shifts complexity from the transmitter (Tx) to the receiver (Rx) to efficiently transmit compressed frequency-domain representations to the back-end hardware for downstream inference processing. While MIPI~\cite{MIPI} already provides a high-speed and energy-efficient interface (up to 14 Gbps with CSI-3), conventional PHY implementations rely on inverter-based Tx drivers and dynamic-comparator-based Rx detection~\cite{Razavi2015TheSL}, which require relatively high signal swing at the transmitter to maintain low bit error rate (BER). To address this, we modify the Rx architecture by introducing a dynamic integrator before the comparator~\cite{Int_Rx}. In this integrating receiver (IR), the input signal is accumulated over the bit period, converting constant logic levels into ramps with positive or negative slopes depending on the transmitted bit. This temporal integration improves the effective signal-to-noise ratio (SNR), allowing reliable detection even with reduced input amplitude. As a result, for a given BER target, the Tx signal swing—and hence Tx power—can be significantly reduced. The proposed design uses fully dynamic (clocked) integrator and comparator circuits, eliminating static power consumption. Due to precharge and evaluation phases, the effective integration time is half the bit period; however, this naturally enables dual data rate (DDR) operation~\cite{Int_Rx}, maintaining high throughput. Overall, this asymmetric architecture enables substantial energy savings at the sensor-side transmitter while preserving robust high-speed communication, and can be readily integrated with FPGA or custom backends for downstream processing.

\section{Experimental Results}

\subsection{Experimental Setup}
\label{exp_set}
We evaluate on CIFAR-10~\cite{CIFAR-10}, Tiny-ImageNet-200~\cite{hansen2015tiny}, and Visual Wake Words (VWW)~\cite{chowdhery2019visual} for image classification, and on COCO~\cite{lin2014microsoft} for object detection and instance segmentation. All images are resized to $224{\times}224$.

We pair our DCT tokenizer with several pretrained backbones: ViT-Tiny/16, ViT-Base/16, Swin-Tiny~\cite{liu2021swin}, and EfficientFormer-L1~\cite{li2022efficientformer}. 
Each serves as both a standalone baseline and the backbone for its FrequencyFormer variant, in which the standard patch embedding is replaced by our multi-scale DCT tokenizer. We evaluate two tokenizer sizes: the small (24-channel) tokenizer selects ($k_Y$, $k_{Cb}$, $k_{Cr}$) = (14, 5, 5) coefficients in Branch~1 and Branch~3; the large (48-channel) selects ($k_Y$, $k_{Cb}$, $k_{Cr}$) = (28,10,10). Branch~2 always extracts (96,24,24) coefficients and adjusts only the output projection dimension by a convolutional layer to match channel size. 
For the $14{\times}14$ high-resolution mode (Tiny-ImageNet and VWW), Branch~1 replaces its strided convolution with $k{=}2, s{=}2$; Branches~2 and~3 remain unchanged. 
When HAQ is applied, the transition harmonic index is set to $t{=}4$ for the $8{\times}8$ block DCT (Branch~1), $t{=}12$ for the $32{\times}32$ block DCT (Branch~2), and $t{=}56$ for the full-image DCT (Branch~3). The minimum precision $b_{\min}$ is set to INT4.

All backbone weights are initialised from ImageNet-21k pretrained checkpoints (ImageNet-1k for EfficientFormer-L1); the DCT tokenizer is randomly initialised. For classification, we train with AdamW~\cite{loshchilov2017decoupled} (lr $1\mathrm{e}{-4}$, weight decay $1\mathrm{e}{-3}$, batch size 512) using cosine annealing with a 10-epoch warm-up, for 200 epochs on CIFAR-10 and Tiny-ImageNet and 100 epochs on VWW. For detection and segmentation, we integrate proposed FrequencyFormer tokenizer into ViTDet~\cite{li2022exploring} with Mask R-CNN~\cite{he2017mask} architecture and train for 20 epochs with AdamW (lr $1\mathrm{e}{-6}$). Standard random-resized cropping and horizontal flipping are applied during training; validation uses a centre crop. For FrequencyFormer models, images are converted to YCbCr for the DCT, while the frozen vanilla patch embedding layer receives ImageNet-normalised RGB. For the T-KD loss, we sweep $\lambda \in \{1, 10, 100\}$ and select the best value per dataset.





\subsection{Main Results}
\FloatBarrier
\begin{table}[!h]
    \centering
    \fontsize{6.5pt}{5pt}\selectfont
    \setlength{\tabcolsep}{2.5pt}
    \caption{
\footnotesize Classification accuracy (\%) on CIFAR-10, Tiny-ImageNet, and VWW. ``INT$n$'' denotes uniform $n$-bit quantization; ``INT$n$-HAQ'' applies Harmonic-Aware Quantization with linear bit-width scheduling from INT$n$ down to INT4.
}
    \vspace{-3mm}
    \label{tab:cls}
    \begin{tabular}{ll |ccc| ccc |ccc}
        \toprule
          &  & \multicolumn{3}{c}{\textbf{CIFAR-10}} & \multicolumn{3}{c}{\textbf{Tiny-IN}} & \multicolumn{3}{c}{\textbf{VWW}} \\
        \cmidrule(lr){3-5}
        \cmidrule(lr){6-8}
        \cmidrule(lr){9-11}
        \multicolumn{2}{l|}{\textbf{Configs}} &ViT-Ti &EF-l1 &Sw-Ti &ViT-B &EF-l1 &Sw-Ti &ViT-Ti &EF-l1 &Sw-Ti \\
        \midrule
        \multicolumn{2}{l|}{Vanilla, FP16}
            & 95.23 &94.98  &96.94
            & 84.75 &85.19 &84.61
            & 91.19 &91.03 &92.45 \\
        \midrule
        \multirow{8}{*}{\rotatebox{90}{
        \shortstack{\textit{48-channel}}
        }}
        & INT9, incl.\ DCT & 94.01 &94.28 &96.31 & 83.21 &83.89 &83.08 &  89.43 &90.51 &91.34  \\
        & INT8, incl.\ DCT & 93.77 &94.01 &96.11 & 82.56 &83.69 &82.17 &  89.21 &90.23 &90.99 \\
        & INT7, incl.\ DCT & 93.32 &93.24 &96.28 & 82.39 &82.95 &80.96 &  88.23 &89.89 &90.54  \\
        & INT6, incl.\ DCT & 92.44 &92.58 &93.88 & 82.41 &80.17 &79.56 &  88.13 &89.11 &89.86  \\
        & INT9-HAQ  & 93.91&93.89 & 96.11 & 83.20  &83.70 &83.14 &89.29 &90.31 &91.17\\
        & INT8-HAQ  & 93.72&93.77 &96.24 & 82.62  &83.72 &82.18 &89.16 &89.73 &90.74  \\
        & INT7-HAQ  & 93.30&93.10 &96.22 &  82.29 &82.56 &81.12 &88.12 &89.12 & 90.41\\
        & INT6-HAQ      & 92.83&91.63 &93.68 &  81.79 &80.02 &79.34  &87.81 &88.89 &88.83 \\
        \hline
        \multirow{8}{*}{\rotatebox{90}{
        \shortstack{\textit{24-channel}}
        }}
        & INT9, incl.\ DCT  &93.24  &93.11  & 95.97 &82.11  &82.37  &82.76  &88.79 &89.58  &90.64 \\
        & INT8, incl.\ DCT  &92.85  &92.55  &95.58 &82.01  &81.96  &82.02  &88.38 &89.87  &90.22\\
        & INT7, incl.\ DCT  &92.44  &92.21  &94.14 &81.16  &80.47  &81.01  &88.23 &89.08  &89.16 \\
        & INT6, incl.\ DCT  &91.78  &91.91  &93.79 &79.02  &79.69  &79.22  &87.38 & 88.39 &88.32\\
        & INT9-HAQ   &93.20  &93.23  &95.71  &82.18&82.44&82.79 &88.59&89.20&90.41   \\
        & INT8-HAQ   & 92.89 &92.37  &95.54  &81.79&81.79&81.74 &88.35&89.66&90.02 \\
        & INT7-HAQ   & 92.42 &92.13  &94.45  &81.24&80.41&79.78 &88.01&89.03&88.93 \\
        & INT6-HAQ   & 91.93 &91.67  &93.45  &79.11&79.50&79.21 &87.80&88.45&88.16 \\
        \bottomrule

    \end{tabular}
    \vspace{-4mm}
\end{table}

\begin{table}[!h]
\vspace{-1mm}
  \small
  \caption{Comparison with SOTA In-/Near-sensor work.}
  \label{tab:sota}
  \vspace{-3mm}
  \centering
  {\fontsize{7pt}{5pt}\selectfont
  \setlength{\tabcolsep}{4pt}
  \begin{tabular}{lc|cccc}
    \toprule
    \textbf{Method}  & \textbf{Task} & \textbf{Res.}  & \textbf{Network}  &\textbf{TOPS/W} & \textbf{Acc.} \\
    \midrule
    Senputing~\cite{xu2021senputing}   & MNIST& 28$^2$ & 2-layer MLP & 4.7& 93.76\\
    SCAMP~\cite{bose2020fully} & MNIST& 256$^2$ &2-layer CNN & 0.535 & 93.0 \\
    MR-PIPA~\cite{abedin2022mr} & MNIST & 256$^2$     &  3-layer CNN & 1.89& 97.26\\
    DPCE~\cite{xu2020utilizing}       &CIFAR10& 32$^2$ & LeNet-5 &11.49  &87.20 \\
    PIPSIM~\cite{roohi2023pipsim}     &CIFAR10&64$^2$ & LeNet-5 &4.12 & 90.05 \\
    P$^2$M~\cite{datta2022processing} &VWW &224$^2$ & MobileNetV2 &0.4  &84.3\\
    Ours (24ch, INT6-HAQ)&CIFAR10 &224$^2$ &ViT-Ti &\textbf{28.8} &\textbf{91.93}\\
    Ours (24ch, INT6-HAQ)&VWW &224$^2$ &ViT-Ti &\textbf{28.8}  &\textbf{87.80}\\

    \bottomrule
  \end{tabular}
  }
  \vspace{-3mm}
\end{table}

\subsubsection{Image Classification.}
 \label{sec:res}

Table~\ref{tab:cls} reports classification accuracy across three datasets and quantization configurations. Accuracy degrades gracefully under quantization (93.77\% at INT8, 92.44\% at INT6), and HAQ consistently stays within 0.4\% of uniform quantization, confirming that reducing precision on high-frequency harmonics introduces negligible accuracy loss. Swin-Tiny, EfficientFormer-L1, and results on Tiny-ImageNet and VWW follow consistent trends. The 24-channel tokenizer further halves transmitted coefficients at modest cost comparing to 48-channel(e.g., 92.89\% vs.\ 93.72\% at INT8-HAQ on CIFAR-10).


\subsubsection{Comparison with Prior In-/Near-Sensor Work.}
Table~\ref{tab:sota} compares FrequencyFormer against prior in-/near-sensor approaches across accuracy and energy efficiency (TOPS/W). Prior methods either operate at low resolution with shallow networks—Senputing~\cite{xu2021senputing} and SCAMP~\cite{bose2020fully} on MNIST with 2-layer MLPs/CNNs, DPCE~\cite{xu2020utilizing} and PIPSIM~\cite{roohi2023pipsim} on CIFAR-10 with LeNet-5—or scale to full resolution but with limited accuracy, as P$^2$M~\cite{datta2022processing} achieves only 84.3\% on VWW with MobileNetV2 at 0.4 TOPS/W.

FrequencyFormer surpasses all prior methods in both accuracy and energy efficiency. With the 24-channel INT6-HAQ configuration, it achieves 91.93\% on CIFAR-10 and 87.80\% on VWW at \textbf{28.8 TOPS/W}, exceeding the best prior CIFAR-10 accuracy (PIPSIM, 90.05\%) by 1.88\% and the best prior energy efficiency (DPCE, 11.49 TOPS/W) by 2.5$\times$. 
Notably, FrequencyFormer is the first in-/near-sensor approach to deploy a vision transformer backbone at full $224{\times}224$ resolution while simultaneously achieving state-of-the-art accuracy and energy efficiency under integer-only quantization.

\subsubsection{Object Detection and Instance Segmentation.}
Table~\ref{tab:coco_ap} evaluates FrequencyFormer on COCO using Mask R-CNN with ViTDet. The 48-channel INT8 variant with HAQ achieves 30.18 box AP and 27.02 mask AP, closely matching the vanilla baseline (30.35 / 27.12), confirming that spatial precision is well preserved through the frequency-domain pipeline. The 24-channel configuration incurs a larger drop (27.32 / 26.37), as dense prediction relies more heavily on fine-grained spatial detail. Nevertheless, the 48-channel variant demonstrates that FrequencyFormer can serve as a drop-in backbone for detection and segmentation without architectural modification.

\begin{table}[t]
\centering
\fontsize{8pt}{5pt}\selectfont
\setlength{\tabcolsep}{4pt}
\caption{Average Precision (AP) metrics on the COCO validation set using Mask R-CNN with vanilla ViTDet and proposed FrequencyFormer backbone.}
  \vspace{-4mm}
\label{tab:coco_ap}
\resizebox{\linewidth}{!}{
\begin{tabular}{l|cccccc}
\toprule
&\multicolumn{3}{c}{\textbf{\textit{Det.}}} &\multicolumn{3}{c}{\textbf{\textit{Seg.}}}  \\
\cmidrule(lr){2-4}
\cmidrule(lr){5-7}
\textbf{Backbone}
& AP & AP$_{50}$ & AP$_{75}$ 
& AP & AP$_{50}$ & AP$_{75}$\\
\midrule

ViTDet          & \textbf{30.35 }& 46.98 & \textbf{32.24}  & 27.12 & \textbf{44.17} & 28.26\\
FrequencyFormer (48ch, INT8)    &30.27&\textbf{47.21}&31.78  &\textbf{27.22}&42.42 &\textbf{28.58}  \\
FrequencyFormer (24ch, INT8)    &27.89&44.93&29.50 &26.10 &41.37 &27.19   \\
FrequencyFormer (48ch, INT8-HAQ)    &30.18&46.71&32.17&27.02&44.38&28.11  \\
FrequencyFormer (24ch, INT8-HAQ)    & 27.32&44.61&28.89&26.37&41.53&27.12 \\

\bottomrule
\end{tabular}
}
\vspace{-4mm}
\end{table}

\subsubsection{Ablation Study.}

\noindent\textit{Frequency-domain vs.\ spatial tokenization.}
Table~\ref{tab:ablation}(a) compares FrequencyFormer against a vanilla ViT-Tiny (both in FP16) patch embedding compressed to the same bottleneck shape ($7{\times}7{\times}24/48$) via strided convolution, with T-KD applied in both cases. FrequencyFormer outperforms the spatial bottleneck by 4.37\% (24ch) and 3.13\% (48ch), demonstrating that the DCT basis preserves substantially more task-relevant information than learned spatial downsampling at equivalent compression ratios.

\noindent\textit{Multi-scale branches.}
Table~\ref{tab:ablation}(b) isolates the contribution of each frequency branch on CIFAR-10 (INT8-HAQ). Branch~1 alone yields 90.27\% / 91.03\% (24ch / 48ch). Adding Branch~2 raises accuracy to 91.41\% / 93.29\%, and incorporating Branch~3 further improves it to 92.85\% / 93.77\%.

\noindent\textit{Channel selection strategy.}
Table~\ref{tab:ablation}(c) compares zigzag ordering against Gumbel-Softmax selection (INT8-HAQ applied). Gumbel-Softmax yields modest accuracy gains (+0.18\% for 24ch, +0.10\% for 48ch), but produces non-contiguous coefficient indices that prevent the row/column pruning (Section~3.1.2), increasing DCT FLOPs by 16\% (54.9M vs.\ 47.2M for 24ch). We therefore adopt zigzag ordering as the default, trading a marginal accuracy gap for substantial computational savings.

\begin{table}[t]
\centering
\fontsize{7pt}{5pt}\selectfont
\setlength{\tabcolsep}{1pt}
\caption{Ablation studies on CIFAR-10.}
\vspace{-4mm}
\label{tab:ablation}
\begin{minipage}[t]{0.25\linewidth}
  \centering
  \textbf{(a) Tokenizer}\\[2pt]
  \begin{tabular}{l|cc}
    \toprule
    Method& 24ch & 48ch \\
    \midrule
    ViT-Ti & 88.89 & 91.04 \\
    FF & \textbf{93.26} & \textbf{94.17} \\
    \bottomrule
  \end{tabular}
\end{minipage}%
\hfill%
\begin{minipage}[t]{0.33\linewidth}
  \centering
  \textbf{(b) Multi-scale branches}\\[2pt]
  \begin{tabular}{ccc|cc}
    \toprule
    Br.1 & Br.2 & Br.3 & 24ch & 48ch \\
    \midrule
    \checkmark & & & 90.27 & 91.03 \\
    \checkmark & \checkmark & & 91.41 & 93.29 \\
    \checkmark & \checkmark & \checkmark & \textbf{92.85} & \textbf{93.77} \\
    \bottomrule
  \end{tabular}
\end{minipage}%
\hfill%
\begin{minipage}[t]{0.42\linewidth}
  \centering
  \textbf{(c) Channel selection}\\[2pt]
  \begin{tabular}{l|cc|cc}
    \toprule
    & \multicolumn{2}{c|}{24ch} & \multicolumn{2}{c}{48ch} \\
    \cmidrule(lr){2-3}\cmidrule(lr){4-5}
    Method& Acc. & FLOPs & Acc. & FLOPs \\
    \midrule
    Gumbel & \textbf{93.03} & 54.9M & \textbf{93.87} & 55.2M \\
    Zigzag & 92.85 & \textbf{47.2M} & 93.77 & \textbf{47.8M} \\
    \bottomrule
  \end{tabular}
\end{minipage}
\vspace{-4mm}
\end{table}

\subsubsection{LUT Area and Energy Analysis}
\begin{figure}[!t]
\begin{subfigure}[h]{0.495\linewidth}
\centering
    \includegraphics[width=\linewidth]{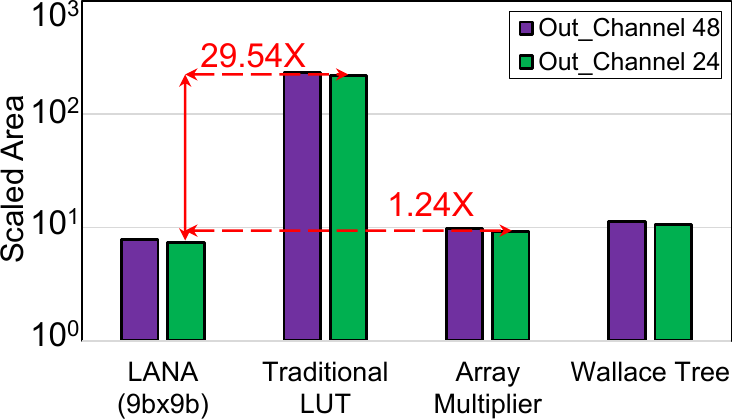}
    \caption{}
\label{fig:areaLANA}
\end{subfigure}
\begin{subfigure}[h]{0.495\linewidth}
\centering
    \includegraphics[width=\linewidth]{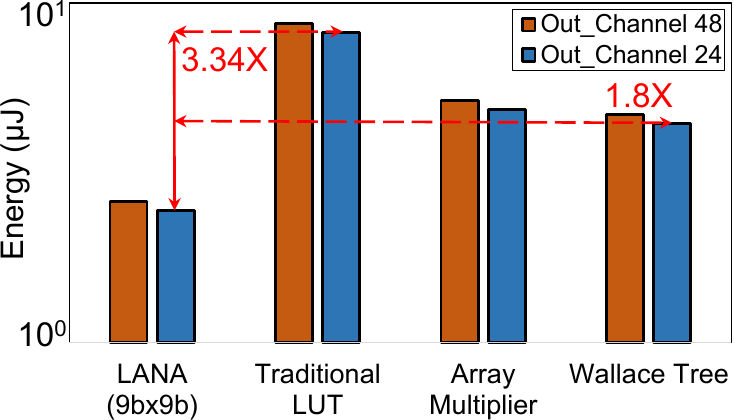}
    \caption{}
\label{fig:energyLANA}
\end{subfigure}
\vspace{-4mm}
\caption{\footnotesize (a) Area overhead for uniform INT9 tokenizer multiplication showing our LUT implementation significantly reduces area compared to Traditional LUT ($\approx$29.54$\times$) and Array Multiplier ($\approx$1.24$\times$). 
(b) Energy per inference comparison, where our LUT achieves up to $\approx$3.34$\times$ and $\approx$1.8$\times$ energy savings over Traditional LUT and Wallace Tree, respectively, demonstrating improved efficiency.}
\label{fig:areaandenergyLANA}
\vspace{-8mm}
\end{figure}

Fig.~\ref{fig:areaandenergyLANA}(a)–(b) evaluate the proposed LUT architecture for uniform INT9 tokenizer multiplication. Hardware overhead is derived from prior LUT-based designs~\cite{sen2025look,dehghanzadeh2025luna}. Our LUT implementation significantly reduces area by avoiding costly multiplier trees and large lookup structures, and lowers energy per inference through reduced computation and switching activity. Even without mixed precision, it offers a strong area–energy baseline. 
Compared to uniform INT9, INT9-HAQ achieves additional reductions of up to 45.9\% and 42.7\% in area (for the 24-channel and 48-channel configurations, respectively), and 40.9\% and 45.9\% in energy, by leveraging smaller LUTs, simpler datapaths, and reduced computation and memory activity.
These gains come with minimal accuracy loss ($<1\%$–$2\%$) across CIFAR-10, Tiny ImageNet, and VWW, indicating that much of the hardware cost of uniform-precision DCT multiplication can be eliminated without degrading performance. Overall, these results show that our method provides an efficient tokenizer multiplication framework across both uniform-precision and mixed-precision modes. The INT9 design already offers strong area and energy benefits, while HAQ variants scaling further enhances these savings with negligible accuracy loss, making the approach highly attractive for energy- and area-constrained edge intelligence systems.

\begin{figure}[!t]

\begin{subfigure}[t]{0.495\linewidth}
\captionsetup{}
\centering
    \includegraphics[width=\linewidth]{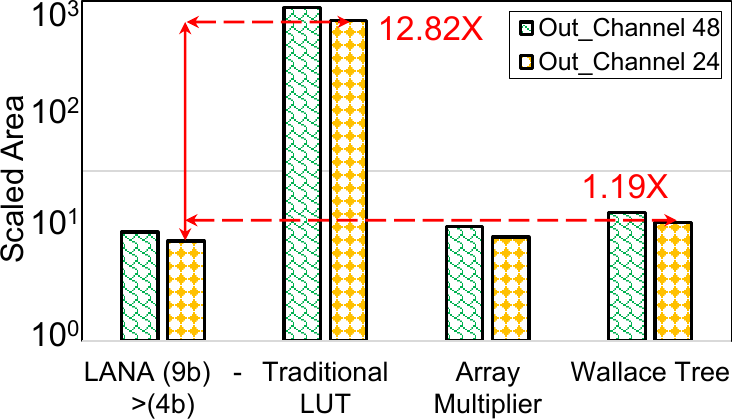}
    \caption{}
\label{fig:areamixedLANApruned}
\end{subfigure}
\begin{subfigure}[t]{0.495\linewidth}
\captionsetup{justification=centering}
\centering
    \includegraphics[width=\linewidth]{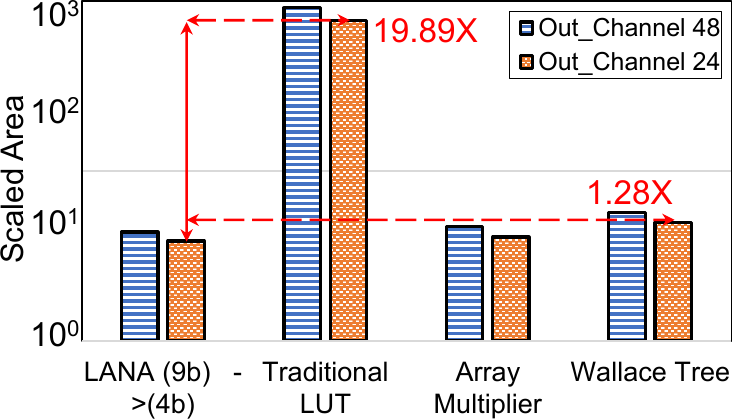}
    
    \caption{\footnotesize{}}
\label{fig:energymixedLANApruned}
\end{subfigure}
\par\smallskip
\begin{subfigure}[h]{0.495\linewidth}
\captionsetup{justification=centering}
\centering
    \includegraphics[width=\linewidth]{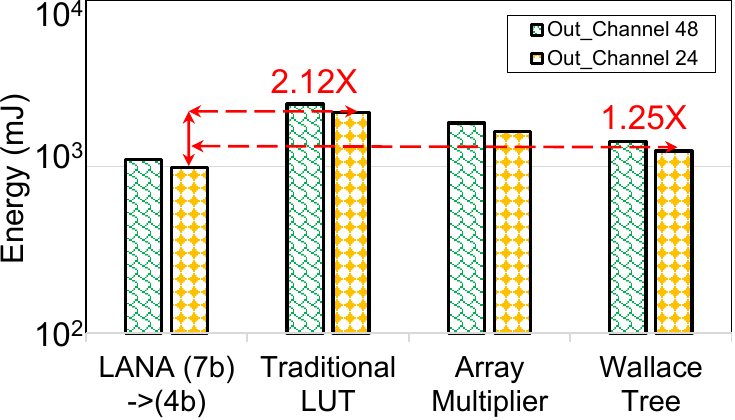}
    \caption{\footnotesize{}}
\label{fig:areamixedLANAtiny}
\end{subfigure}
\begin{subfigure}[h]{0.495\linewidth}
\captionsetup{justification=centering}
\centering
    \includegraphics[width=\linewidth]{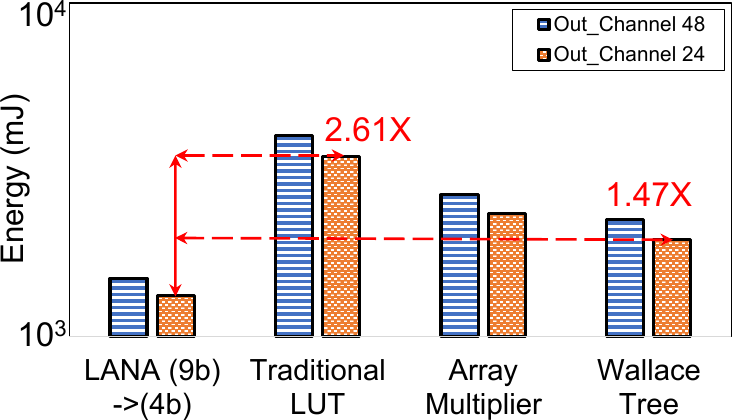}
    \caption{\footnotesize{}}
\label{fig:energymixedLANAtiny}
\end{subfigure}
\vspace{-5mm}
\caption{\footnotesize (a) Area overhead for INT7-HAQ tokenizer multiplication, where our LUT implementation achieves significant area reduction over Traditional LUT ($\approx$12.82$\times$) and Array Multiplier ($\approx$1.19$\times$). 
(b) Area overhead for INT9-HAQ configuration, showing up to $\approx$19.89$\times$ and $\approx$1.28$\times$ area savings over Traditional LUT and Wallace Tree, respectively. 
(c) Energy per inference for INT7-HAQ, demonstrating $\approx$2.12$\times$ and $\approx$1.25$\times$ energy savings over Traditional LUT and Wallace Tree. 
(d) Energy per inference for INT9-HAQ, where our LUT implementation achieves up to $\approx$2.61$\times$ and $\approx$1.47$\times$ energy reduction compared to Traditional LUT and Wallace Tree, respectively.}

\label{fig:areaandenergymixedLANA}
\vspace{-4mm}
\end{figure}

\subsubsection{Sensor Communication Energy Reduction}
\begin{figure}[t]
  \centering
\includegraphics[width=\linewidth]{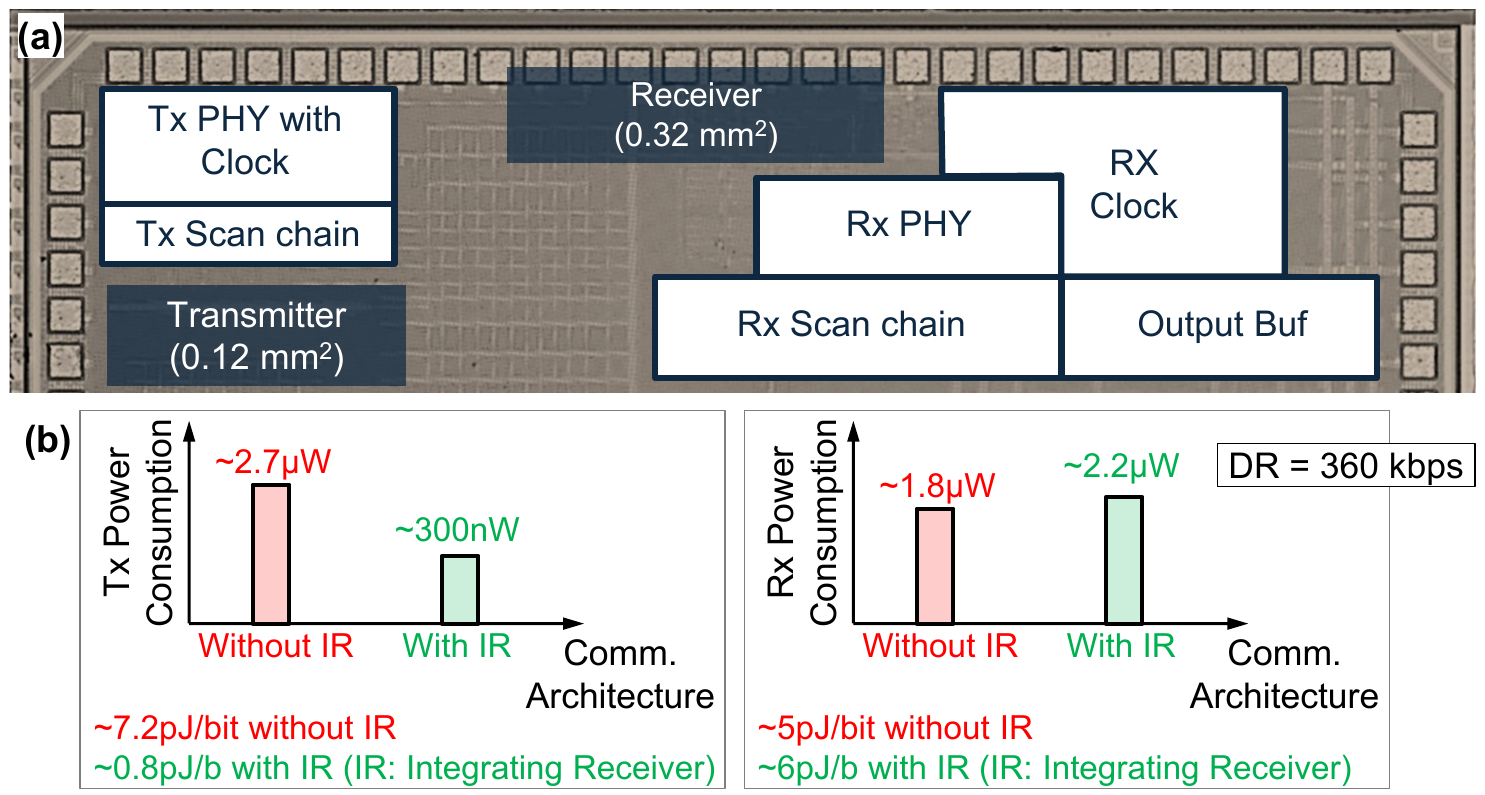}
  \vspace{-4mm}
\caption{\footnotesize (a) 65nm test-chip micrograph for the communication PHY; (b) Power consumption at the Tx and Rx with IR (integrator+comparator) and without IR (comparator only).}
  \label{fig:Comm_results}
  \vspace{-4mm}
\end{figure}

To compute the communication energy at the sensor for the two scenarios (1. comparator only - i.e. without IR, and 2. integrator+comparator, i.e. IR), we demonstrate measured results from a 65nm test chip for same SNR at the input of the comparator. At 30 fps frame rate ($\approx$36 Mbps raw data rate, which reduces to $<$360 kbps through the FrequencyFormer processing pipeline), we have $\approx$ 2.77/2 = 1.38 $\mu$s time to perform the integration. We set the integrator's time-domain gain (configurable in the test chip) in a way that the SNR at the comparator input is $>$ 20 dB (for BER $<$10$^{-4}$) for both scenarios (with and without integrator). This allows for aggressive voltage scaling at the Tx PHY ($\approx$ 0.4V) with IR, whereas for the conventional scenario (without IR), we use 1.2V supply as per standard MIPI operations. With this, the Tx driver's energy efficiency is found to be $\approx$ 0.8pJ/bit with IR, and $\approx$ 7.2pJ/bit without IR (baseline). The Rx front-end energy efficiency is $\approx$ 6pJ/bit with IR, and $\approx$ 5pJ/bit without IR (baseline), with clocking power included in both. Fig. \ref{fig:Comm_results} shows the chip micrograph and power consumption near 360 kbps data rates.

\subsubsection{System-Level Energy Analysis}
\label{sec:system_energy}

We analyze the sensor-side energy budget, covering all costs incurred before tokens enter the backbone: sensor readout, tokenizer computation, and sensor-to-processor communication. The backbone inference cost is excluded as it is identical across all configurations—FrequencyFormer acts as a drop-in replacement for the patch embedding and does not alter the transformer blocks. Table~\ref{tab:system_energy} reports the breakdown across six configurations that progressively activate each contribution.

The baseline system transmits a raw $224{\times}224{\times}3$ image (1,204,224 bits at INT8) over a standard MIPI-CSI link to the processor that executes the full ViT-Tiny backbone. 
In-sensor compute energy for the digital MAC baseline is estimated using 45\,nm synthesis results for array multipliers~\cite{sen2025look}, while our LUT estimates use the D\&C architecture from Section~3.2. Communication energy uses the measured Tx and Rx efficiencies from Section~4.4: for the standard MIPI link, $E_{\mathrm{Tx}}{=}7.2$\,pJ/bit and $E_{\mathrm{Rx}}{=}5.0$\,pJ/bit (total 12.2\,pJ/bit); for the proposed asymmetric link with integrating receiver, $E_{\mathrm{Tx}}{=}0.8$\,pJ/bit and $E_{\mathrm{Rx}}{=}6.0$\,pJ/bit (total 6.8\,pJ/bit). All FrequencyFormer configurations use the 48-channel INT8-HAQ tokenizer unless otherwise noted.

Table~\ref{tab:system_energy} reports the energy breakdown across six system configurations:
\begin{enumerate}
    \item[\textbf{S1}] \textit{Baseline}: raw image $\rightarrow$ standard MIPI $\rightarrow$ processor (ViT-Tiny).
    \item[\textbf{S2}] \textit{Baseline + proposed comm.}: raw image $\rightarrow$ proposed MIPI $\rightarrow$ processor (ViT-Tiny). Isolates the communication interface contribution alone.
    \item[\textbf{S3}] \textit{FF tokenizer (digital MAC) + standard MIPI}: FrequencyFormer tokenizer implemented with conventional array multipliers in-sensor $\rightarrow$ standard MIPI $\rightarrow$ processor. Isolates the bandwidth reduction from frequency-domain compression.
    \item[\textbf{S4}] \textit{FF tokenizer (digital MAC) + proposed MIPI}: same as S3 but with the proposed low-power link. Combines bandwidth reduction and communication efficiency.
    \item[\textbf{S5}] \textit{FF tokenizer (LUT) + standard MIPI}: FrequencyFormer tokenizer with D\&C LUT hardware $\rightarrow$ standard MIPI $\rightarrow$ processor. Combines bandwidth reduction and energy-efficient in-sensor compute.
    \item[\textbf{S6}] \textit{Full proposed pipeline}: FrequencyFormer tokenizer with LUT hardware $\rightarrow$ proposed MIPI $\rightarrow$ processor backbone. All three contributions active.
\end{enumerate}

\begin{table}[t]
\centering
\fontsize{7pt}{5pt}\selectfont
\setlength{\tabcolsep}{3pt}
\caption{System-level energy breakdown per frame ($\mu$J) for ViT-Tiny on CIFAR-10 at 30\,fps. ``Sensor'' = in-sensor tokenizer compute; ``Comm.'' = Tx + Rx link energy; . Best values in \textbf{bold}.}
\vspace{-3mm}
\label{tab:system_energy}
\begin{tabular}{cl|ccc|cc}
\toprule
& & \multicolumn{3}{c|}{\textbf{Energy ($\mu$J)}} & & \\
\cmidrule(lr){3-5}
\textbf{ID} & \textbf{Configuration} & Sensor & MAC/LUT & Comm. & \textbf{Total} & \textbf{Reduction} \\
\midrule
S1 & Baseline (raw + std.\ MIPI) 
   & 9.57 & ---  & 14.69  &24.26 & 1.00$\times$ \\
S2 & Raw + proposed MIPI 
   & 9.57 & --- & 8.19  &17.76 &  1.37$\times$ \\
\midrule
S3 & FF (MAC) + std.\ MIPI 
   & 9.57 & 2.22 & 0.115 & 11.91 & 2.04$\times$ \\
S4 & FF (MAC) + prop.\ MIPI 
   & 9.57 & 2.22 & 0.064 & 11.85 & 2.05$\times$ \\
S5 & FF (LUT) + std.\ MIPI 
   & 9.57 & 1.3 & 0.115 & 11.00 & 2.21$\times$ \\
S6 & \textbf{Full pipeline} 
   & 9.57 & 1.3 & \textbf{0.064} & \textbf{10.93} & \textbf{2.22}$\times$ \\
\bottomrule
\multicolumn{7}{l}{\footnotesize Comm.\ energy: S1/S3/S5 use 12.2\,pJ/bit; S2/S4/S6 use 6.8\,pJ/bit.} \\
\multicolumn{7}{l}{\footnotesize S1--S2 transmit 1,204,224 bits; S3--S6 transmit 9,408 bits (128$\times$ reduction).}
\end{tabular}
\vspace{-4mm}
\end{table}

The communication energy column reveals the multiplicative interaction between the two communication-facing contributions. Frequency-domain compression alone (S3 vs.\ S1) reduces communication energy by $128\times$ by shrinking the data volume from 1.2\,Mb to 9.4\,kb per frame. The proposed MIPI interface alone (S2 vs.\ S1) achieves a $1.8\times$ reduction by lowering the energy per bit from 12.2 to 6.8\,pJ/bit through aggressive Tx voltage scaling enabled by the integrating receiver. When both are combined (S6 vs.\ S1), the communication energy drops by approximately $230\times$ ($128 \times 1.8$), from 14.69\,$\mu$J to 0.064\,$\mu$J per frame. This demonstrates that the two contributions are complementary: bandwidth reduction makes the low-power interface practical (the lower data rate relaxes timing margins and enables deeper voltage scaling), while the interface optimization extracts additional savings from the already-compressed stream. On the sensor side, comparing S5 against S3 isolates the benefit of LUT-based computation over conventional array multipliers for the tokenizer. The LUT realization achieves 1.71$\times$ lower sensor compute energy, consistent with the per-MAC savings reported in Fig.~\ref{fig:areaandenergyLANA}(b) and amplified by the HAQ precision schedule that reduces the effective bit-width (and hence LUT size) for high-frequency harmonics. 

The energy breakdown in Table~\ref{tab:system_energy} shows that communication dominates the baseline (S1, 65\% of total), but becomes negligible after frequency-domain compression (S3--S6, $<$1\%). Table~\ref{tab:system_energy_ablation} disaggregates the per-component contributions: frequency-domain compression provides the largest single reduction (128$\times$ in communication), LUT-based hardware lowers the cost of in-sensor tokenization (1.71$\times$ over array multipliers), and the low-power MIPI interface provides multiplicative savings on the already-compressed stream. Together, the full pipeline (S6) achieves 2.22$\times$ total energy reduction over the baseline,  while maintaining 93.72\% accuracy on CIFAR-10 (within 1.51\% of the vanilla FP16 baseline).

\begin{table}[t]
\centering
\fontsize{7pt}{5pt}\selectfont
\setlength{\tabcolsep}{5pt}
\caption{Per-component energy reduction factors, computed by toggling each contribution independently.}
\label{tab:system_energy_ablation}
\vspace{-3mm}
\begin{tabular}{l|cc}
\toprule
\textbf{Component} & \textbf{Comparison} & \textbf{Reduction} \\
\midrule
Freq.-domain compression (comm.) & S3 vs.\ S1 & $128\times$ \\
Proposed MIPI interface (comm.) & S4 vs.\ S3 & $1.8\times$ \\
Combined comm.\ savings & S6 vs.\ S1 & $\approx230\times$ \\
\midrule
LUT vs.\ digital MAC (sensor) & S5 vs.\ S3 & $1.71\times$ \\
\bottomrule
\end{tabular}
\vspace{-4mm}
\end{table}


\section{Conclusion}
We presented FrequencyFormer, a co-designed sensor-to-processor pipeline that enables practical deployment of vision transformers on bandwidth-constrained edge systems. The pipeline comprises three tightly integrated components: a multi-scale DCT tokenizer that compresses raw images into compact frequency-domain tokens with up to 128$\times$ data reduction, a LUT-based hardware realization that exploits the fixed-coefficient structure of DCT for energy-efficient near-sensor computation, and a low-power MIPI-based communication interface that further reduces energy per bit through aggressive voltage scaling and asymmetric transceiver design. Together, the full pipeline achieves 2.22$\times$ sensor-side energy reduction and $\approx$230$\times$ communication energy reduction over the conventional baseline, while attaining 28.8 TOPS/W—2.5$\times$ higher energy efficiency than SOTA work. By operating as a drop-in replacement for the standard patch embedding, FrequencyFormer preserves compatibility with pretrained ViT backbones and generalizes across classification, detection, and segmentation tasks with minimal accuracy degradation. Our results establish frequency-domain tokenization as a principled and hardware-friendly foundation for pushing vision transformer inference to the point of image capture.

\bibliographystyle{ACM-Reference-Format}
\bibliography{references}

\end{document}